\documentclass[11pt,a4paper]{article}
\usepackage{jheppub}
\pdfoutput=1
\usepackage{bm, bbm, slashed}
\usepackage{subcaption}

\usepackage{chngcntr}

\DeclareMathOperator{\real}{Re}

\title{Evolution of Holographic Fermi Arcs from a Mott Insulator}
\author{Garrett Vanacore,}
\author{Srinidhi T. Ramamurthy,}
\author{Philip W. Phillips\thanks{Guggenheim Fellow}
}
\affiliation{Department of Physics and Institute for Condensed Matter Theory \\ University of Illinois, 1110 W. Green Street, Urbana, IL 61801}

\emailAdd{dimer@illinois.edu}

\arxivnumber{1506.06769}

\abstract{We study fermions in an electrically-probed and asymptotically anti-de Sitter Schwarzschild spacetime which interact via novel chiral symmetry-preserving interactions. Computing the dual fermion two-point correlator, we show that these bulk interactions anisotropically gap Fermi surfaces of the boundary spectrum. Consequently, the interactions we devise provide holographic models for Fermi arcs seen ubiquitously in the pseudogap regime of the cuprates. Our interactions are modifications of the chiral symmetry-breaking Pauli coupling, which has previously been proposed as the holographic realization of Mott physics. The onset of Mott insulation and pseudogap physics are respectively discussed in the context of bulk chiral and boundary parity symmetry breaking, and the Mott transition is interpreted as a deconfinement transition of non-Fermi liquid excitations.}

\notoc

\begin{document}
\counterwithout{equation}{section}
\maketitle
\flushbottom
\newpage


Superconductivity in the copper-oxide ceramics remains unresolved largely because of the unconventional electronic properties of the normal state.
For example, when holes are doped into the copper-oxide planes, the metallic state that ensues is not characterized by a continuous closed surface in momentum space as dictated by Landau's paradigmatic theory of metals.  Rather, the surface is truncated, forming what are referred to as Fermi arcs \cite{norman98,johnson,shen,KingP11,he}.    This stark deviation from the standard theory of metals can be viewed in one of two ways: either some type of order \cite{chakravarty,norman} gives rise to a Fermi pocket with momentum-dependent spectral intensity that is vanishingly small for some range of momenta, or the problem is inherently rooted in strong coupling physics in which zeros of the single-particle electronic Green function, caused by a divergent self-energy \cite{tsvelik,konik2006,yrz,shong,kotstan},  are at the base of the vanishing spectral weight.  These scenarios are distinguished based on their adherence to the Luttinger sum rule \cite{kdave}. 
Within the former, quasiparticles exist but carry spectral weight too small to be measured experimentally on the `back half' of the Fermi pocket, and so Luttinger's rule is satisfied.   In the latter, however, whenever $\det \real G(\omega=0,\mathbf{k})=0$ the Luttinger rule is inapplicable \cite{kdave,rosch,altshuler}.  Experimentally, the measured Fermi surface areas \cite{johnson,shen,he} at zero magnetic field violate Luttinger's rule, a state of affairs which persists even at large fields \cite{qoscill}. The central problem of Fermi arcs thus appears to be elucidating how strong interactions persist from the Mott insulating state and partially gap the Fermi surface of the doped state.

An ideal resolution of this problem would utilize a non-perturbative method to account for the strong interactions in the Mott state while delineating a mechanism for Fermi arc formation.  While Fermi arcs have been obtained numerically \cite{kotstan} and phenomenologically \cite{yrz}, they have evaded analytical methods that are valid in the strongly coupled regime. To address this shortcoming, we utilize the gauge/gravity duality \cite{maldacena,witten,gubser} (or `holography') --- a method relating the physics of strongly interacting quantum systems to that of weakly interacting gravitational systems in one higher dimension --- to investigate Fermi arc formation from non-Fermi liquid states. Prior holographic studies have generated Fermi arcs by anisotropically condensing fermions into p-wave \cite{vegh2010} or d-wave \cite{herzog2011} superconducting states, but this cannot describe cuprate physics because the cuprate arcs form at temperatures above the superconducting transition. In contrast, we present a mechanism for arc formation which does not utilize superconductivity, and therefore represents a distinctly different state. The key result of this work is quite clear. States in which a gap forms without manifest symmetry breaking, hereafter referred to as Mott states, are realized at the boundary of asymptotically anti-de Sitter geometries by bulk fermions undergoing chiral symmetry-breaking interactions. In contrast,  Fermi arcs obtain from the breaking of a discrete symmetry at the boundary, induced by chiral symmetry-preserving interactions in the bulk. We provide concrete illustrations of the latter in bottom-up models with $(2+1)$-dimensional and parity-broken boundary duals, and argue that boundary theories of chiral symmetry-invariant holographic fermions are most naturally interpreted as two-fluid models undergoing momentum space confinement/deconfinement transitions.  

While there are several ways \cite{sslee,faulkner,zaanen,gubser,gauntlett} to implement the holographic program for fermion matter at finite density, we pursue a bottom-up construction in which the  action for a bulk $(d+1)$-dimensional gravitational system is supplemented with fermionic fields that source operators at the $d$-dimensional boundary.  Since this procedure provides only correlation functions for the boundary theory, there is considerable lee-way in choosing the fermion interactions in the bulk. Schematically our action
	\begin{align} \label{eq:Lbulk}
		{S}_{\text{bulk}} = {S}_{\text{grav}} + {S}_{\text{gauge}} + {S}_{\text{fer}}
	\end{align}
will consist of gauge and gravity sectors, $S_{\rm gauge}$ and $S_{\rm grav}$ respectively, with $S_{\rm fer}$ describing probe fermionic fields which source fermion operators in the boundary conformal theory.  The only restriction on the gauge and gravitational parts of the action is that they provide geometries which asymptote to anti-de Sitter (AdS) spacetime at the boundary.  In fact, one of the key conclusions of our work is that our results are independent of the detailed gravitational structure of the bulk, provided it contains electromagnetism and some form of a black hole.

Our focus at the outset is the fermionic part of the action. Fermi to non-Fermi liquid behavior has previously been shown \cite{faulkner,zaanen} to emerge from the simple choice of the Dirac action minimally coupled to four-dimensional Reissner-Nordstr\"om-AdS (RN-AdS$_4$), in which bulk fermions $\psi$ of mass $m$ source boundary fermion operators of scaling dimension $d/2 \pm mL$, with $L$ the curvature radius of the asymptotic AdS geometry. The two possible scalings of the boundary operators follow from independent prescriptions for identifying the operators' sources and responses, referred to as standard and alternative quantization (resp.\ $\pm$). This construction also produces gapped spectra as $m$ is increased, but this has not been tied to Mott physics and the model's symmetry forbids descriptions of pseudogaps.
Nonetheless, there are a number of non-minimal gauge interactions that can be added to extend this model, the simplest of which is the Pauli coupling,
 \begin{align} \label{eq:LPauli}
		{S}_{\text{fer}} =\int d^{d+1}x  \sqrt{-g}\ i  \overline{\psi} \left( \slashed{D} - m - i p \slashed{F} \right)\psi,
	\end{align}
with $p$ controlling the strength of a dipole interaction between the fermionic and Maxwell fields. Structurally, the dipole interaction provides shifts in fermion momenta that depend on the the boundary chemical potential; this charge scale shifts only the fermion frequencies in the minimal model. Tuning $p$ from large negative ($<-1$) to large positive ($> 1$) values (in the conventions of Ref.\ \cite{edalati2}) results in diverse phenomenology of the boundary fermions: the dominant low-frequency pole in their spectrum passes through regimes of Fermi liquid-, marginal Fermi liquid-, and non-Fermi liquid-like scaling before reaching a gapped phase. While the gapped structure was thought \cite{edalati1,edalati2} to obtain from a vanishing of the quasiparticle residue, in actuality it arises from an exact pole--zero duality within the diagonal blocks of the boundary fermion propagator,
\begin{align} \label{eq:pzduality}
	G_{ii} (\omega, k; m, p) = -\frac{1}{G_{ii}(\omega, -k; -m, -p)},
\end{align}
first shown for RN-AdS$_4$ \cite{alsup} and later for electrically-probed Schwarzschild-AdS$_4$ (SS-AdS) \cite{garrett}. The inverse relationship \cite{faulkner}  between $G_{ii} (\omega, k; m, p)$ and $G_{ii}(\omega, -k; -m, -p)$, ultimately rooted \cite{garrett} in the two quantizations for holographic fermions,  was not exploited until it was realized \cite{alsup} that the boundary spectrum must solely exhibit zeros for large positive $p$, as it contains only poles at large negative $p$. It is well-known \cite{tsvelik,konik2006,shong,kotstan,kdave,dzy} that the Mott gap requires zeros of the single-particle propagator.  Consequently, the vanishing of the spectral weight is due to zeros and the bulk Pauli coupling in RN-AdS$_4$  and electrically-probed SS-AdS$_4$ mimics Mott physics.  

There is a subtlety, however, in the Mott gap formed from the Pauli interaction. Although the symmetries of the boundary spectrum are preserved, the Pauli term has a non-zero anticommutator with the generator of chiral rotations, $ \{ \Gamma^{ab},\Gamma_5 \} \ne 0$.
That is, chiral symmetry is broken in the bulk.  This is not entirely surprising since chiral symmetry breaking is a typical mechanism for the generation of mass.  However, more than the static breaking of this symmetry is relevant here.   What is crucial  to note is that the Pauli term only generates a gap for sufficiently large and {\it positive} values of $p$; there would be no such restriction if the gap were attributable merely to loss of chiral symmetry. Because the Pauli term changes the scaling dimension of the dual boundary operators and  increasing the exponent  converts pole singularities to zeros, it is ultimately the dynamical breaking of chiral symmetry that is the root cause of the gap. The same reasoning applies to the gap generated by the Dirac mass $m$ in the minimal model, though in that case gapping ensues without alteration of the coupling between bulk fermions and the boundary charge scale. Though the dimensionality of the bulk and boundary differ, they share time coordinates and fermion charges and so must share time reversal and charge conjugation symmetries. Bulk chiral symmetry breaking must then be reflected in discrete symmetries of the boundary. Extrapolating to flat-space lattice models, where chiral symmetry is a combination of particle--hole and time reversal symmetry, we may take the holographic results to mean that the generation of Mott-type gaps requires the breaking of one of these symmetries.

Because the pseudogap is not a completely gapped phase, we expect holographic models for Fermi arcs to preserve bulk chiral symmetry.  Exploiting the pole--zero duality inherent in the Pauli construction, we anticipate that a bulk interaction which couples fermion momenta to different signs of the Pauli term should generate both poles and zeros and hence Fermi arcs.  Consequently, we propose the following bulk fermion action to model  Fermi arcs,
	\begin{align} \label{eq:Sfermiarc}
		S_{\text{fer}} = \int d^4 x \sqrt{-g} \ i  \overline{\psi} \left( \slashed{D} - m - i \wp_1 \Gamma \slashed{F} + \wp_2 (\hat{n} \cdot \vec{\Gamma}) \slashed{F} \right) \psi.
	\end{align}
The non-minimal interactions tuned by $\wp_1$ and $\wp_2$ differ from the dipole interaction of \eqref{eq:LPauli} through the presence of the matrices $\Gamma \equiv \Gamma^{\underline{r}} \Gamma^{\underline{t}} ( \hat{n} \cdot \vec{\Gamma} )$ and $\hat{n} \cdot \vec{\Gamma}$, which restore bulk chiral symmetry while breaking rotational and Lorentz symmetries of the boundary theory. The loss of boundary rotational invariance is necessary to model anisotropic phenomena such as Fermi arcs and is characterized in our model by the breaking of parity along the unit vector $\hat{n}$.\footnote{In two dimensions parity is defined by a sign change of only one spatial coordinate, which requires a choice of reflection line. The vector $\hat{n}$ provides a normal to the line across which boundary parity is broken.} If preserving bulk symmetries is preferred, an interaction like the second can be engineered using a bulk vector field $\chi$ constrained by a boundary condition. Consider the fermion action
	\begin{align} \label{eq:Sfermiarc2}
		S_{\text{fer}} = \int d^4 x \sqrt{-g} \ i  \overline{\psi} \left( \slashed{D} - m + \wp_3 \slashed{\chi} \slashed{F} \right) \psi.
	\end{align}
If $\chi$ is taken as a constant solution of its equation of motion, the reduced form of the action above yields an interaction nearly identical to the $\wp_2$ interaction, save for an extra frame field factor. For simplicity we examine the interactions individually.

Beginning with the $\wp_1$ interaction, we may naively infer the effects of $\Gamma$ from our understanding of the pole--zero duality and the Pauli coupling model. From the work of Refs.\ \cite{edalati2, garrett} we know that, with $m=0$, the Pauli model produces sharp Fermi surfaces at large negative $p$ and gapped spectra at large positive $p$. In terms of the two diagonal entries of the boundary fermion Green function $G_{ij} (\omega, \mathbf{k})$, these Fermi surfaces manifest in two poles: one of $G_{11}$ at $(\omega=0, |\mathbf{k}_F|)$, and one of $G_{22}$ at $(\omega=0, -|\mathbf{k}_F|)$, with the Fermi momentum $\mathbf{k}_F$ depending on the background geometry. The $\wp_1$ interaction enters the bulk Dirac equation in the same way as the Pauli coupling along the $\hat{n}$-momentum axis --- preserving the general pole/zero structure of $G_{11}$ and $G_{22}$ in this frame --- but inverts the sign of $p$ in the $G_{11}$ block. Consequently, in the $\hat{n}$ frame the pole that was once present at $k_F$ becomes a zero at $ -k_F$, coincident with the pole still present in $G_{22}$, and when $G_{22}(0, k_F; p)$ and $G_{11} (0, k_F; -p)$ have negligible spectral weight the Fermi surface gaps at $k_F$. 
Conversely, the $\wp_2$ interaction is a less straightforward modification of the dipole coupling, and therefore does not manifestly engineer pole/zero coexistence in the boundary spectrum.\footnote{The dipole and $\wp_1$ interactions couple the same bulk fermion degrees of freedom in the $\hat{n}$ frame, differing only in the signs of those couplings. The $\wp_2$ interaction couples different degrees of freedom, so the pole/zero structure it induces at the boundary cannot be immediately inferred from the dipole model.} Its virtues are preservation of bulk chiral symmetry and omission of the radial boost generator $\Gamma^{\underline{r}} \Gamma^{\underline{t}}$ present in the $\wp_1$ interaction; the former property is demanded by our line of reasoning, while the latter gives the interaction a form more aesthetically natural from the perspective of the boundary, where radial boosts lack straightforward interpretation.

\begin{figure} 
	\begin{center}
	\includegraphics[scale=.278]{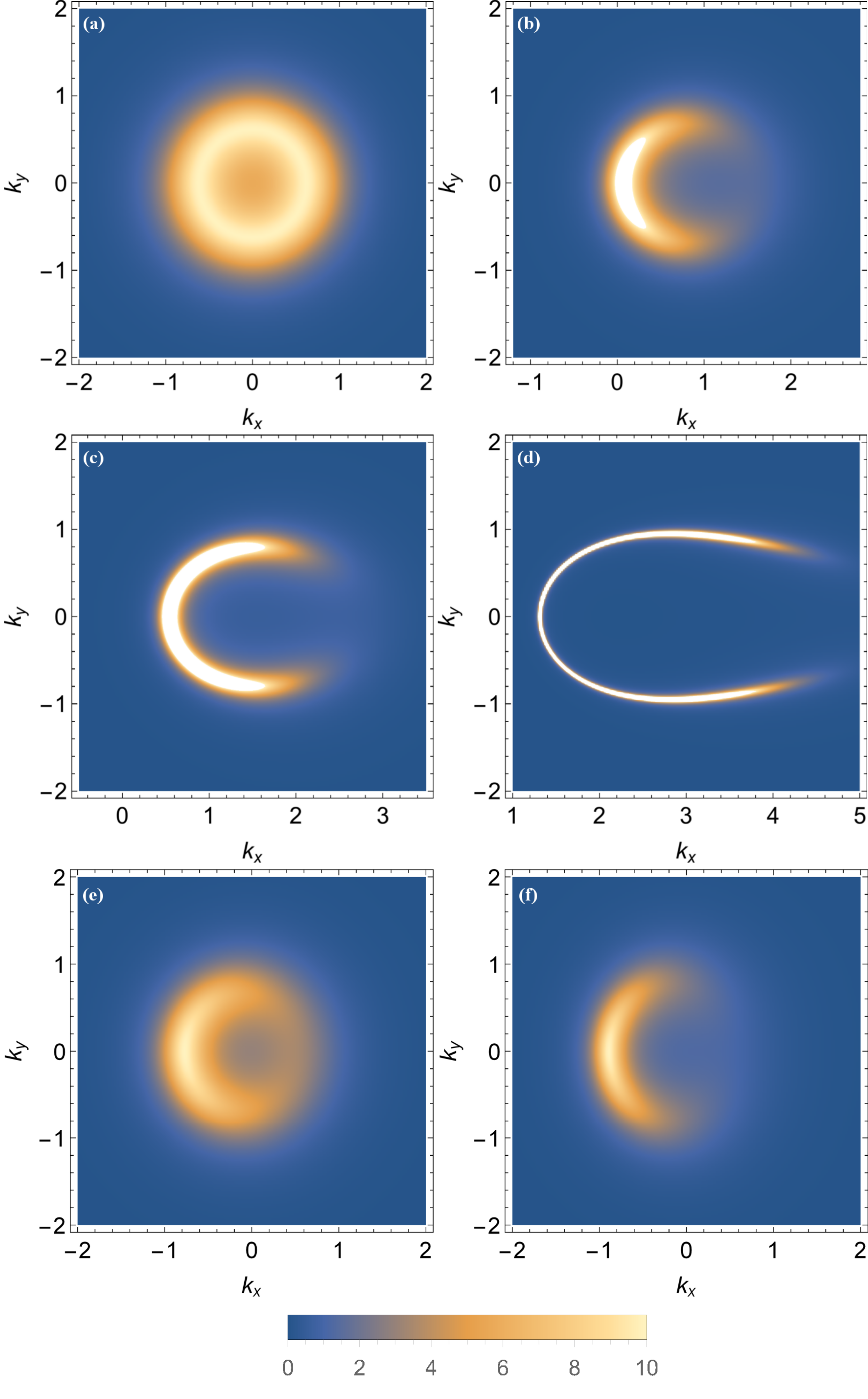}
	\end{center}
	\caption{Boundary fermion spectral functions $A(\omega, k_x, k_y)$, at $\omega = 10^{-3} + i \delta$, $\delta = 10^{-6}$, in the Fermi arc model \eqref{eq:Sfermiarc} with couplings of $\wp_1 = 0$, $1$, $2$, $4$ with $\wp_2 =0$ (panels (a)--(d)), and with $\wp_2 =0.2$, $0.4$ with $\wp_1 = 0$ (panels (e)--(f)). Other bulk parameter choices are listed in the appendix.} \label{fig:spectra}
\end{figure}

\begin{figure} 
	\begin{center}
	\includegraphics[scale=.4]{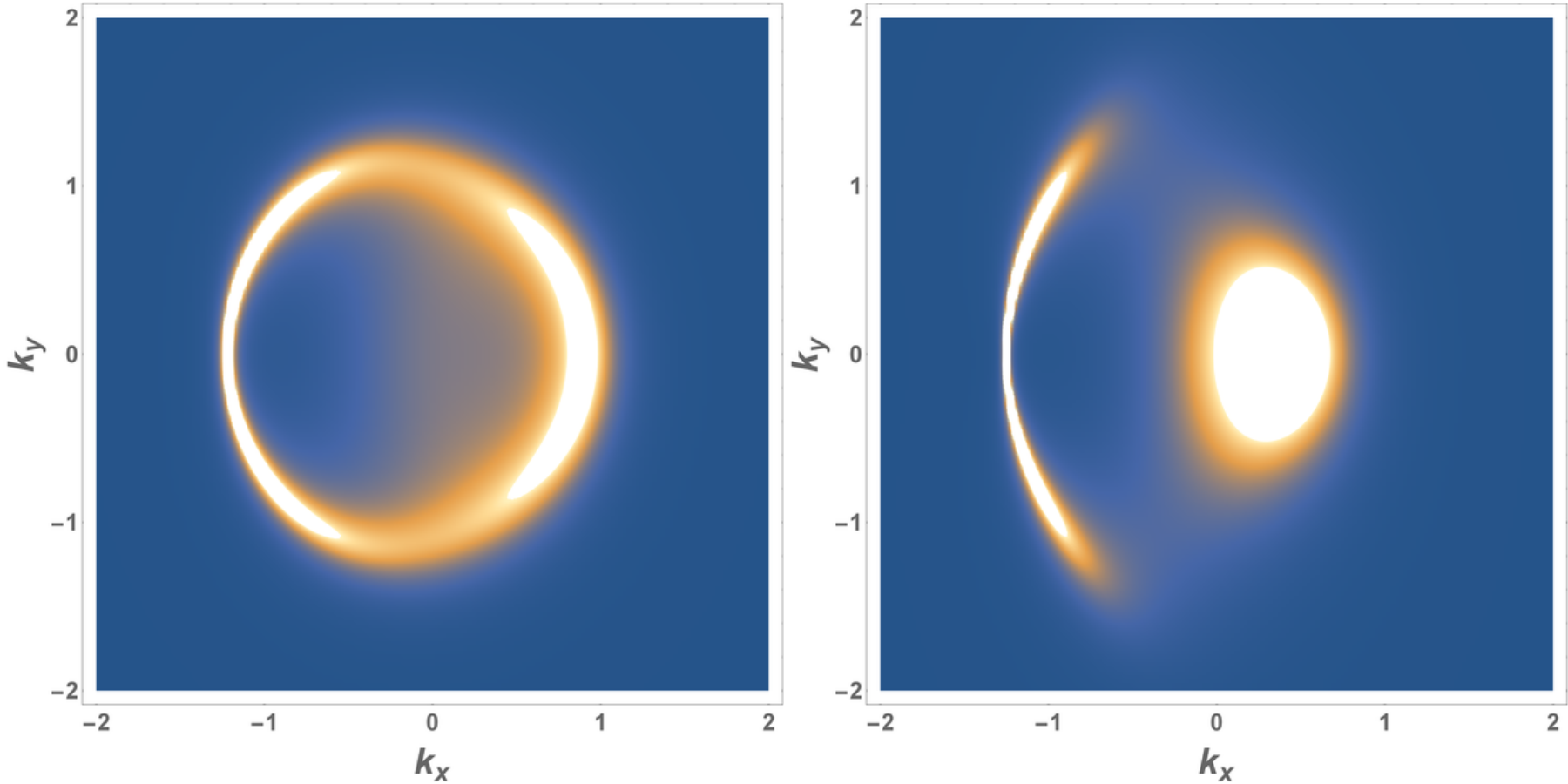}
	\includegraphics[scale=.9]{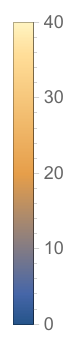}
	\end{center}
	\caption{Boundary fermion spectral functions $A(\omega, k_x, k_y)$, with the same modeling parameters as Fig.\ \ref{fig:spectra}, in the Fermi arc model \eqref{eq:Sfermiarc} with couplings of $\wp_3 = 1$ and $1.8$ (left, right).} \label{fig:spectra2}
\end{figure}

While the above considerations provide some intuition about the spectrum in the $\hat{n}$ frame, the model's lack of rotational symmetry makes it difficult to infer spectral properties at general momenta. To remedy this we have numerically computed the boundary fermion spectral function in planar SS-AdS$_4$ with a Maxwell probe. Qualitatively identical results are found in extremal RN-AdS$_4$. Unnormalized spectral densities of the boundary fermions at the small frequency $\omega = 10^{-3} + i \delta$, for four values of the $\wp_1$ interaction and two of $\wp_2$, both with $\hat{n} = \hat{x}$, are presented in Fig.\ \ref{fig:spectra}. Spectral densities resulting from two values of the $\wp_3$ interaction, with $\chi = dx$, are presented in Fig.\ \ref{fig:spectra2}. A small broadening factor $\delta = 10^{-6}$ was introduced to resolve poles in the retarded Green function. More information about the geometry and computation of Green functions, including parameters used to generate Fig.\ \ref{fig:spectra}, may be found in the appendix. 

In the absence of the non-minimal interactions, electrically-probed SS-AdS$_4$ hosts a highly-broadened Fermi surface in the boundary dual. Once the $\wp_1$ or $\wp_2$ interaction is switched on, the previously-discussed gapping process ensues: spectral weight on the right half of the Fermi surface is suppressed, while that on the left half is enhanced. This trend continues as $\wp_1$ and $\wp_2$ are increased, leading to an unmistakably arc-like spectrum as Fig.\ \ref{fig:spectra} reveals. If either coupling constant is instead tuned to negative values, the spectral suppression and enhancement occur on the left and right halves of the Fermi surface (respectively). The $\wp_3$ interaction yields arcs as well, but of a slightly different variety; gaps form off the $k_x$ axis, while a pocket of spectral weight forms opposite the arc on the $k_x$ axis. Thus, our bulk interactions necessarily provide gapless spectra in the boundary dual, in contrast with the chiral symmetry-breaking interactions that have previously been studied. Note that the center of mass momentum produced by our interactions may be canceled by introducing a second, independent flavor of bulk fermions that experience the same interaction with coupling $-\wp_i$.  Further, our mechanism generates arcs without the aid of superconductivity \cite{vegh2010,herzog2011}, and hence could provide a framework for understanding the emergence of arcs in the cuprates.

From the bulk perspective the arc formation is most easily understood in terms of bulk fermion orbits. Technically, poles and zeros of boundary fermion propagators reflect the linear independence of fermion sources and responses; these objects are identified with the boundary values of the two (two-component) $\Gamma^{\underline{r}}$ eigenspinors, and the propagator is defined as the transformation matrix that relates them (see appendix). Yet the bulk is nothing but an on-shell scattering problem for Dirac particles in a radial potential encoding gravity and electromagnetism.\footnote{Quantum corrections are in principle suppressed through the weak-coupling/strong-coupling duality between bulk and boundary.} Numerical solutions of the Dirac equation (with in-falling boundary conditions) reveal that the $\Gamma^{\underline{r}}$ eigenspinors assume non-normalizable scattering states for most boundary wave modes. At certain frequencies and momenta, however, one of the spinors (or more generally one of its components) can enter a normalizable bound state; this is precisely where poles and zeros appear in the boundary spectral function. Holographic Fermi surfaces are attributed \cite{faulkner} to the `source' spinor occupying a bound state. Similarly, zeros appear at frequencies and momenta where the `response' spinor is bound to the black hole. The relationship between the pole--zero duality and the choice of quantization scheme naturally follows. Our model \eqref{eq:Sfermiarc} thus achieves a coexistence of poles and zeros, and exhibits Fermi arcs, by mediating bound states for both types of fermions via its non-minimal interactions.

Understanding the arc from the boundary perspective is a more difficult task, as the field theory mechanisms behind holographic Fermi surfaces are under debate (competing interpretations are discussed in Refs.\ \cite{dewolfe} and \cite{swingle}), while those behind holographic zeros have not been investigated. We can make some interpretive progress, however, by appealing to holographic generalizations of Luttinger's rule \cite{liuiqbal2012}. These generalizations illustrate how on-shell Dirac fermions --- i.e., bulk charges external to a black hole --- provide a traditional Luttinger count by adding the volumes contained by Fermi surfaces to the boundary charge. In contrast, charges contained by a black hole provide an explicit deviation from the traditional count. But there is a critical oversight in these works as they ignore the possibility of zeros. The Pauli model provides many instances of zero surfaces in the boundary spectrum, and by the arguments here zeros are attributable to charges outside the black hole. Thus, even in the holographic context, zeros represent charged degrees of freedom which lack structure in momentum space. Our Fermi arc model then implies that the pseudogap is a two-fluid state composed of momentum-confined and momentum-deconfined charges. Building a field theoretic understanding of the latter would not only be interesting in its own right, but may also elucidate how pseudogaps intervene the development of non-Fermi liquids from Mott insulating states. 

Given the central role played by gravity and electromagnetism in the bulk, it is natural to wonder how backreaction of fermions will affect results from probe models.  The key questions concern how  bulk geometries and fermion orbits are altered by backreaction. Fermion orbits are observed widely enough in bottom-up and top-down models for us to expect their presence in generic AdS-black hole geometries, but it is not obvious that backreacted systems will support the $\omega=0$ bound states necessary for boundary zeros and Fermi surfaces.  What is encouraging, however, is that recent work on the existence of Fermi surfaces in top-down constructions \cite{gubser2015} can be understood entirely in terms of a competition between gravity and electromagnetism.   In the cases studied \cite{gubser2015}, Fermi surfaces are always observed when there are bulk fermions with (positive) charge appreciably greater than their mass, provided the black hole hosts the corresponding electric field. In marginal cases where fermion charges and masses are comparable, the existence of a Fermi surface is contingent upon the presence of a positive Pauli interaction.  Hence, investigating backreacted geometries should reveal whether or not the presence of zeros leads to an increase in the charge behind the horizon as a result of  infall of fermion bound states.   Such infall would be consistent with Mott insulation arising from deconfined charges behind the horizon.

\appendix
\subsection*{Appendix: Model details and calculation of correlators}
We study bulk fermions with non-minimal interactions, described by the Lagrangian
	\begin{align} \label{eq:Lfermiarc}
		{\cal L}_{\text{fer}} = i \sqrt{-g} \  \overline{\psi} \left( \slashed{D} - m - i \wp_1 \Gamma \slashed{F} + \wp_2 (\hat{n} \cdot \vec{\Gamma}) \slashed{F} \right) \psi,
	\end{align}
with $\Gamma \equiv \Gamma^{\underline{r}} \Gamma^{\underline{t}} (\hat{n} \cdot \vec{\Gamma})$ and $\vec{\Gamma} \equiv (\Gamma^{\underline{x}}, \Gamma^{\underline{y}})$. We take $\hat{n} = \hat{x}$ henceforth. The covariant derivative and Maxwell tensor may be written
	\begin{align}
	\begin{split}
		\slashed{D} &= e^M_c \Gamma^c \left( \partial_M + \frac{1}{4} \omega_M^{ab} \Gamma_{ab} - i q A_M \right), \\
		\slashed{F} &= \frac{1}{2}  \Gamma^{ab}  e^M_a e^N_b F_{MN},
	\end{split}
	\end{align}
with $e^M_a$ the (inverse) vielbein, $\omega_M^{ab}$ the spin connection, and $\Gamma^{ab} \equiv \frac{1}{2} [\Gamma^a, \Gamma^b]$. Our index conventions use capital Roman letters for bulk coordinates $M, N \cdots = \{ t, x^i, r \}$ and lower-case Roman letters for tangent space coordinates $a,b \cdots = \{ \underline{t}, \underline{x}^i, \underline{r} \}$.

For the geometric background we choose Schwarzschild-AdS in $(d+1) = 4$ dimensions. Parametrizing in the Poincar\'e patch, the line element may be written
	\begin{align} \label{eq:metric}
		ds^2 = \frac{r^2}{L^2} \left( - f(r) dt^2 + d\mathbf{x}^2 \right) + \frac{L^2}{r^2} \frac{dr^2}{f(r)}.
	\end{align}
We take the black hole to have unit (dimensionless) mass, so the emblackening factor is given by
	\begin{align}
		f(r) = 1 - \left( \frac{r_0}{r} \right)^3,
	\end{align}
with $r_0$ denoting the horizon radius. The temperature of the boundary theory is then $T = 3 r_0/4 \pi L^2$. We give the Maxwell probe the form familiar from RN-AdS$_4$,
	\begin{align} \label{eq:maxwell}
		A = \mu \left(1 - \frac{r_0}{r} \right) dt, \quad \mu = Q r_0 / L^2,
	\end{align}
and treat $Q$ as a tuning parameter for the boundary chemical potential $\mu$. When we refer to electrically-probed SS-AdS in the main text, we mean the metric \eqref{eq:metric} with the probe Maxwell field \eqref{eq:maxwell}. The plots in Fig.\ \ref{fig:spectra} were generated with $Q = \sqrt{3}$, $r_0 = L =  q =1$, and $m=0$.

To evaluate the Dirac operator it is convenient to Fourier transform the bulk spinor in the boundary coordinates and scale it by the factor $r^{3/2} f^{1/4}$ to eliminate the spin connection. Hence we take $\psi (r, x) \sim \psi(r,k) r^{3/2} f^{1/4} e^{i k \cdot x}$, with $k \equiv (\omega, \mathbf{k})$. Choosing the following basis for the Dirac matrices,
	\begin{align}
	\label{eq:diracmatrices}
	\begin{split}
		\Gamma^{\underline{r}} &=
			\begin{pmatrix}
				- \sigma_3  & 0 \\
				0 & - \sigma_3
			\end{pmatrix}, \quad
		\Gamma^{\underline{t}} = 
			\begin{pmatrix}
				i \sigma_1& 0 \\
				0 & i \sigma_1
			\end{pmatrix}, \\
		\Gamma^{\underline{1}} &= 
			\begin{pmatrix}
				- \sigma_2   & 0 \\
				0 &  \sigma_2
			\end{pmatrix}, \quad
		\Gamma^{\underline{2}} = 
			\begin{pmatrix}
				0 & \sigma_2  \\
				\sigma_2  & 0
			\end{pmatrix},
	\end{split}
	\end{align}
and rescaling the non-minimal couplings as $\wp_i \rightarrow \wp_i L$, the Dirac equation for the scaled fields yields two coupled equations,
\begin{align}
	\label{eq:diraceqn_fermiarcs}
	\begin{split}
		\frac{r^2}{L^2} \sqrt{f(r)} \partial_r \psi_{j} =& \frac{i \sigma_2}{\sqrt{f(r)}} \left( \omega + q \mu \left(1- \frac{r_0}{r} \right) \right) \psi_{j} - \sigma_3 \left(\frac{r}{L} m + (-1)^j \wp_2 \mu \frac{r_0}{r} \right) \psi_{j} \\
		&\quad- (-1)^j \sigma_1 \left( \wp_1 \mu \frac{r_0}{r} - k_1 \right) \psi_{j} + \sigma_1 k_2 \psi_{i}.
	\end{split}
\end{align}
Here we have expanded the spinor as $\psi = ( \psi_1, \psi_2)^T$, with the two-component spinors $\psi_{j}$ reflecting the block structure of \eqref{eq:diracmatrices}. The final factor of $\psi_i$ should strictly take $i\neq j$. For a more detailed derivation, see Ref. \cite{edalati2}.

The retarded Green functions of boundary fermion operators are realized by asymptotic solutions to the bulk Dirac equation, subject to in-falling conditions on the bulk spinors at the black hole horizon. Asymptotically the solutions to \eqref{eq:diraceqn_fermiarcs} are $\psi_{j} (r, k) = (b_{j} (k) r^{-mL}, a_{j} (k) r^{mL} )^T$, with $j \in \{1, 2\}$. In the mass window $|mL| < 1/2$ we are free to choose the $a_j$ or the $b_j$ as the sources for fermion operators in the dual theory. Though there is no distinction between these ``quantization" schemes in the zero mass case of the minimal model, we make the standard choice of $A= (a_{1},a_2)^T$ as the source and $B= (b_{1},b_2)^T$ as the vacuum expectation value of a boundary fermion operator. Generally the source and vev are related through a linear transformation, $B = {\cal S} A$, from which the Green may be computed via \cite{liuiqbal2009}
	\begin{align} \label{eq:greenfunc}
		G (\omega, \mathbf{k}) = -i {\cal S} \gamma^{\underline{t}},
	\end{align}
with $\gamma^{\underline{t}}$ a boundary Dirac matrix ($\gamma^{\underline{t}} = i \sigma_1$ in our basis). The causal structure of this Green function is determined by boundary conditions on the bulk spinors at the black hole horizon, with in-falling conditions providing the data for retarded correlators.

Though it is possible in principle to compute the transformation matrix ${\cal S}$ through numerical integration of the Dirac equation, in practice it is often easier to work with a set of first-order, non-linear evolution equations for the Green function components. To implement this procedure, we first expand the Dirac equation about the horizon to determine a basis of in-falling states for the bulk spinors. The near-horizon expansion gives
	\begin{align}
		\frac{(r-r_0)}{ \tilde{\omega}} \partial_r
			\begin{pmatrix}
				\psi_1 \\
				\psi_2
			\end{pmatrix}
		=
			\begin{pmatrix}
				i \sigma_2 & 0 \\
				0 & i \sigma_2
			\end{pmatrix}
			\begin{pmatrix}
				\psi_1 \\
				\psi_2
			\end{pmatrix},
	\end{align}
where $\tilde{\omega} \equiv \omega L^2/ r_0 d$. Writing the two-component bulk spinors as $\psi_j = (\beta_j, \alpha_j)^T$, in-falling solutions $\xi^{\rm I}$, $\xi^{\rm II}$ are given by the eigenvectors of $\text{diag}(i \sigma_2, i \sigma_2)$ with eigenvalue $-i$ , or
	\begin{subequations}  \label{eq:infalling}
	\begin{align}
		\xi^{\rm I} &=
			\begin{pmatrix}
				i ,1 , 0, 0
			\end{pmatrix}^T,
			 \quad  \beta_1^{\rm I} = i, \ \alpha_1^{\rm I} = 1,\  \beta_2^{\rm I} = \alpha_2^{\rm I} = 0, \\
		\xi^{\rm II} &=
			\begin{pmatrix}
				0, 0, i, 1
			\end{pmatrix}^T,
			\quad \beta_1^{\rm II} = \alpha_1^{\rm II} = 0, \  \beta_2^{\rm II} = i, \  \alpha_2^{\rm II} = 1.
	\end{align}
	\end{subequations}
The components $\alpha_j$ and $\beta_j$ act respectively as sources and responses for fermion operators at the conformal boundary, with the Green function given by \eqref{eq:greenfunc}. The data from both sets of boundary conditions can be encoded in a single equation by defining the matrices \cite{faulkner2010}
\begin{align}
		Y =
			\begin{pmatrix}
				\beta_1^{\rm I} & \beta_1^{\rm II} \\
				\beta_2^{\rm I} & \beta_2^{\rm II}
			\end{pmatrix}, \quad
		Z =
			\begin{pmatrix}
				\alpha_1^{\rm I} & \alpha_1^{\rm II} \\
				\alpha_2^{\rm I} & \alpha_2^{\rm II}
			\end{pmatrix}, \quad
		Y = G Z,
\end{align}
such that $G$ asymptotically realizes the boundary Green function \eqref{eq:greenfunc}. Taking a derivative of the third equality and making use of the Dirac equation \eqref{eq:diraceqn_fermiarcs}, we derive the following evolution equation for $G$,
	\begin{align}
		\frac{r^2}{L^2} \sqrt{f(r)} \partial_r G &= M_+ + GM + MG -G M_- G \label{eq:floweqn},
	\end{align}
where the matrices $M_{\pm}$ and $M$ are
	\begin{align}
	\begin{split}
		M_{\pm} &=
			\begin{pmatrix}
				\pm v_{\pm} (r) - k_x & k_y \\
				k_y & \pm v_{\mp} (r) + k_x
			\end{pmatrix}, \\
		M &= \frac{1}{r}
			\begin{pmatrix}
				\wp_2 \mu r_0  & 0 \\
				0 & -\wp_2 \mu r_0
			\end{pmatrix},
	\end{split}
	\end{align}
and
	\begin{align}
		v_{\pm} (r) \equiv \frac{1}{\sqrt{f(r)}} \left( \omega + q A_{t} (r) \right) \pm \wp_1 \mu \frac{r_0}{r}.
	\end{align}
The in-falling conditions \eqref{eq:infalling} provide the initial condition
	\begin{align}
		\lim_{r \rightarrow r_0} G =
			\begin{pmatrix}
				i & 0 \\
				0 & i
			\end{pmatrix},
	\end{align}
which allows numerical integration of \eqref{eq:floweqn}.

\acknowledgments

We thank Rob Leigh, Tom Faulkner, and Onkar Parrikar for their characteristically level-headed remarks and the NSF DMR-1461952 for partial funding of this project. STR acknowledges support from the ONR YIP Award N00014-15-1-2383. PP also acknowledges partial support from the Center for Emergent Superconductivity, a DOE Energy Frontier Research Center, Grant No.\ DE-AC0298CH1088 and the J.\ S.\ Guggenheim Foundation.

\bibliographystyle{JHEP}
\bibliography{holo_fermiarcs}

\end{document}